\documentclass[journal]{IEEEtran}

\usepackage{amsmath}
\usepackage{amssymb}
\usepackage{amsfonts}
\usepackage{amsthm}
\usepackage{enumitem}
\usepackage[ruled]{algorithm2e}
\usepackage{algpseudocode}
\usepackage{comment}
\usepackage{epsfig}
\usepackage{graphicx}
\usepackage{color}
\usepackage{cite}
\usepackage{nomencl}
\usepackage{xstring}
\usepackage{xstring}
\usepackage{xpatch}
\usepackage[utf8]{inputenc} 
\usepackage{graphicx} 
\usepackage{cleveref} 
\usepackage{multirow}

\title{On the Learning of Digital Self-Interference Cancellation in Full-Duplex Radios}
\author{Jungyeon~Kim, Hyowon~Lee, Heedong~Do, Jinseok~Choi,\\ Jeonghun~Park, Wonjae~Shin, Yonina~C.~Eldar, and Namyoon~Lee}

\begin{document}

\maketitle

\begin{abstract}
Full-duplex communication systems have the potential to achieve significantly higher data rates and lower latency compared to their half-duplex counterparts. This advantage stems from their ability to transmit and receive data simultaneously. However, to enable successful full-duplex operation, the primary challenge lies in accurately eliminating strong self-interference (SI). Overcoming this challenge involves addressing various issues, including the nonlinearity of power amplifiers, the time-varying nature of the SI channel, and the non-stationary transmit data distribution. In this article, we present a  review of recent advancements in digital self-interference cancellation (SIC) algorithms. Our focus is on comparing the effectiveness of adaptable model-based SIC methods with their model-free counterparts that leverage data-driven machine learning techniques. Through our comparison study under practical scenarios, we demonstrate that the model-based SIC approach offers a more robust solution to the time-varying SI channel and the non-stationary transmission, achieving optimal SIC performance in terms of the convergence rate while maintaining low computational complexity. To validate our findings, we conduct experiments using a software-defined radio testbed that conforms to the IEEE 802.11a standards. The experimental results demonstrate the robustness of the model-based SIC methods, providing practical evidence of their effectiveness. 
\end{abstract}

\section{Introduction}
Inband full-duplex communications have been widely considered as a potential solution for increasing spectral efficiency in future wireless networks. The fundamental idea behind this approach is to simultaneously transmit and receive data signals at the same frequency \cite{bharadia2013full, Sabharwal2014, Chung2015Proto, Liao2015cognitive, barento2021Joint, Ian2023Spatial}. In principle, this approach has the potential to double the spectral efficiency without requiring additional bandwidth. Moreover, when combined with advanced scheduling algorithms, it can also improve the cell and network throughput of unlicensed-band communication systems using listen-before-talk protocols.  

In practice, however, realizing the potential performance gains is not without challenges, as it is required to cancel a high self-interference (SI) signal at a receiver. In inband full-duplex communications, the transmit signal gives rise to very strong interference to the received signal of interest, which makes SI cancellation (SIC) more complicated. For example, the SI signal strength is about 100 dB larger than the desired signal power \cite{bharadia2013full}. Intuitively, SIC can be achieved by subtracting the transmit from the received signal using the knowledge of the transmit signal waveform. In practice, SI signal cancellation is very challenging because the transmit signal experiences several radio frequency (RF) components such as filters, oscillators, and power amplifiers (PA), which result in linear and non-linear distortions of the transmit signal. In addition, the reflectors surrounding the mobile transceiver can generate time-varying SI signals. Thus, accurate cancellation of the time-varying and nonlinear SI signal is indispensable to enable inband full-duplex radios for next-generation wireless systems. 

\begin{figure*}
\centering
\includegraphics[width=0.9\linewidth]{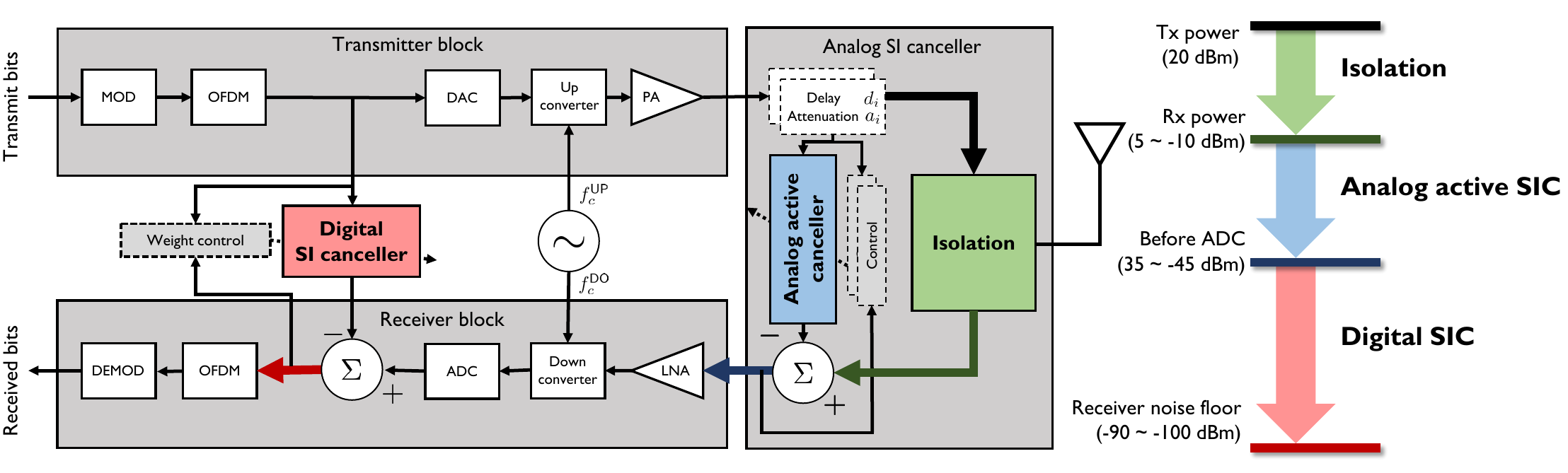}
\caption{A block diagram of a full-duplex wireless system cascaded by circulator isolation, analog SIC, and digital SIC blocks.}
\label{fig: Model}
\end{figure*}

In practical implementation, a SIC block in full-duplex radios typically involves a cascaded approach, utilizing an analog passive SI canceller, an analog active SI canceller, and a digital nonlinear adaptive SI canceller, as shown in Fig. \ref{fig: Model}. \cite{bharadia2013full, Sabharwal2014, Chung2015Proto}. The analog passive SI canceller such as isolation and analog active SI canceller play a crucial role in attenuating the strong self-interference power to a level below the saturation threshold of the analog-to-digital converter (ADC), preventing signal saturation. On the other hand, the digital SIC is designed to further reduce the residual self-interference after analog SIC to the noise power level. Unlike analog SIC, digital SIC needs to address both the nonlinearity effects of the power amplifier (PA) and the time-varying channel response, which renders the digital SIC implementation challenging\cite{bharadia2013full, Korpi2014Widely, Korpi2016Fullduplex,  kim2018adaptive, kong2022neural}.
 
The nonlinear SIC problem is mathematically equivalent to nonlinear system identification in adaptive filter theory \cite{haykin2002adaptive}. Therefore, traditional digital SIC algorithms are primarily based on mathematical models  designed using domain knowledge \cite{bharadia2013full, Korpi2014Widely,kong2022neural, Alexios2018neural, attar2022parallel}. The key merits of these model-based approaches are their computational efficiency and mathematical interpretability. However, these model-based methods may fail to represent the nonlinear SI channel when model-mismatch effects are pronounced. With the advent of machine learning, there has been a shift towards using data-driven methods, such as kernel-based adaptive filters \cite{attar2022parallel} and deep neural networks (DNN) \cite{kong2022neural, Alexios2018neural}. These model-free techniques learn the complex SI channel through data-driven methods, utilizing sets of transmit and receive data samples. However, this approach is susceptible to changes in system environments and is less explainable compared to model-based approaches.

The question of whether to use modeling or not for nonlinear and time-varying digital SIC is a crucial consideration in implementing SIC algorithms for full-duplex radios. However, despite its significance, most prior studies fail to adequately compare and address this particular topic \cite{bharadia2013full, Korpi2014Widely,kong2022neural, Alexios2018neural,  attar2022parallel}. In this article, we aim to address this question by highlighting the trade-offs of model-based and data-driven digital SIC algorithms for full-duplex radios.

The article is structured as follows. In Section \ref{sec2}, we provide an overview of state-of-the-art model-based digital SIC algorithms, focusing on how they utilize models to capture nonlinear distortions in the SI channel. We compare four different model-based digital SIC algorithms, highlighting their strengths and weaknesses in terms of SIC performance and implementation costs. In Section \ref{sec3}, we present model-free approaches, which include kernel and DNN-aided SIC algorithms that do not rely on explicit modeling of the SI channel. In Section \ref{sec4}, we compare the performance of model-based and model-free SIC algorithms. In Section \ref{sec5}, we evaluate a prototype implementation of the SIC algorithms using a software-defined radio (SDR) testbed compliant with the IEEE 802.11a Wi-Fi standards. Finally, our conclusion is given with a discussion in Section \ref{sec7}.

\begin{figure*}
\centering
\includegraphics[width=0.9\linewidth]{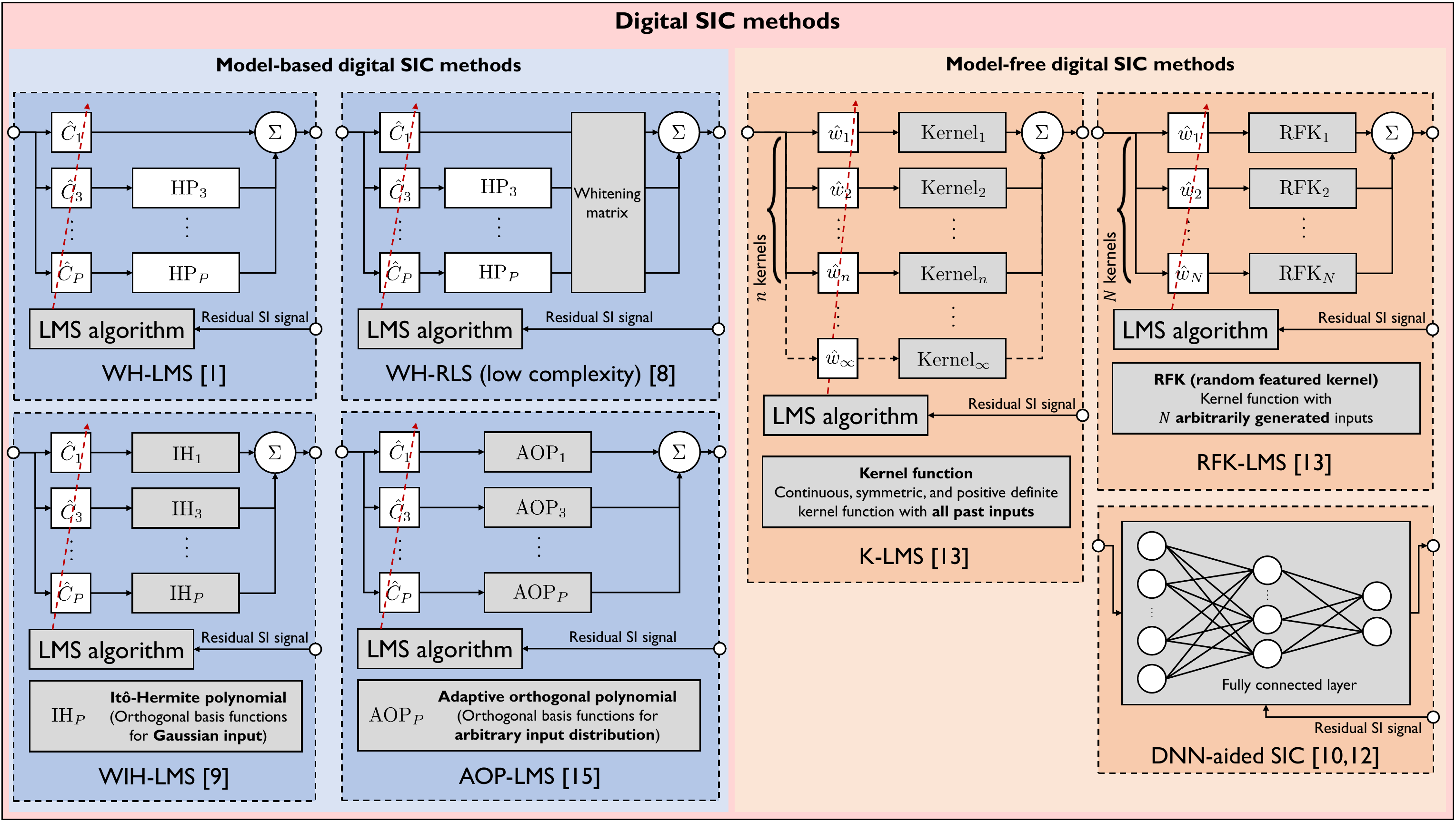}
\caption{Block diagrams of the model-based and the model-free digital SIC algorithms.}
\label{fig: SIC}
\end{figure*}
 
\section{Model-Based Digital SIC} \label{sec2}
Model-based SIC algorithms leverage mathematical formulations to represent the complex characteristics of nonlinear and time-varying SI channels. These are commonly based on domain knowledge such as the input-output relationship of the power amplifier (PA) and the time-varying channel impulse response.

There are two important questions when developing model-based SIC algorithms.
\begin{itemize}
    \item How to construct models that represent the SI channel accurately with a small number of model parameters?
    \item How to optimize the model parameters efficiently?
\end{itemize}
The first question is about modeling accuracy and complexity. The latter one is about algorithm efficiency. Depending on the modeling and optimization methods, existing SIC algorithms can be categorized into four different types, which will be explained in the sequel and Fig. \ref{fig: SIC}. 

\subsection{WH Model with LMS Optimizer }
One widely used approach is harnessing the Wiener-Hammerstein (WH) model \cite{ding2004digital}, which consists of parallel Hammerstein polynomials (HPs) and linear finite impulse response (FIR) filters. This model offers notable advantages for representing SI channels by using domain knowledge. 
  
The HPs in the WH model map the transmit signal into multiple nonlinear signals using HP basis functions, allowing for flexible parameterization of the nonlinearity of the PA. Moreover, increasing the order of HPs enhances the model's ability to capture the nonlinear characteristics of the SI channel. The FIR filters in the model capture the dynamics of the linear transfer function in the SI channel, and increasing the number of FIR filter taps provides higher degrees of freedom for adapting to the dynamics of the linear transfer function. Another advantage of the WH model is its transparent relationship with linear systems, making it easier to implement in practical SIC algorithms. In contrast, other nonlinear models such as kernel-based and neural network models often require more complex parameterization.
 
The WH model is utilized to parameterize the SI channel with parallel FIR filter coefficients for a given number of HPs. Consequently, the model parameters, which are the coefficients of the FIR filters, need to be optimized using a set of transmit and receive data samples as training data. The widely used objective for optimizing these parameters is to minimize the mean squared- error (MSE) between the desired and actual signals. In full-duplex radios, online sample-by-sample optimization is necessary to effectively track the dynamics of the SI channel. The least mean squares (LMS) algorithm, a stochastic gradient descent method, is commonly employed for SIC due to its simple implementation \cite{bharadia2013full}.

\subsection{WH Model with RLS Optimizer}

The main limitation of using the WH model in conjunction with LMS algorithms for SIC is the slow convergence rate, resulting from statistical correlations among the nonlinearly transformed signals by the HP basis functions. To address this low rate convergence issue, it is crucial to eliminate the correlation among the transformed signals. One approach is to use the recursive least squares (RLS) algorithm, which adaptively finds FIR filter coefficients to minimize the mean squared error, and offers faster convergence compared to LMS. However, RLS has extremely high computational complexity due to the estimation and inversion of the covariance matrix of transformed signals in an online fashion.

Recently, a low-complexity variant of RLS for digital SIC has been proposed in \cite{Korpi2016Fullduplex}, where the covariance matrix is estimated in an offline fashion and its inverse is used for orthogonalizing the transformed signals. Despite the faster convergence speed, this low-complexity RLS approach still requires large system memory to store the covariance matrix, which can be a limitation. Additionally, it is limited to use only when the transmit data symbols are stationary, as it cannot accurately orthogonalize transformed signals in the presence of non-stationary data signals resulting from adaptive modulation and coding techniques. This inaccurate covariance information can degrade the performance of SIC.
 
\begin{table*}
\centering
\includegraphics[width=0.9\linewidth]{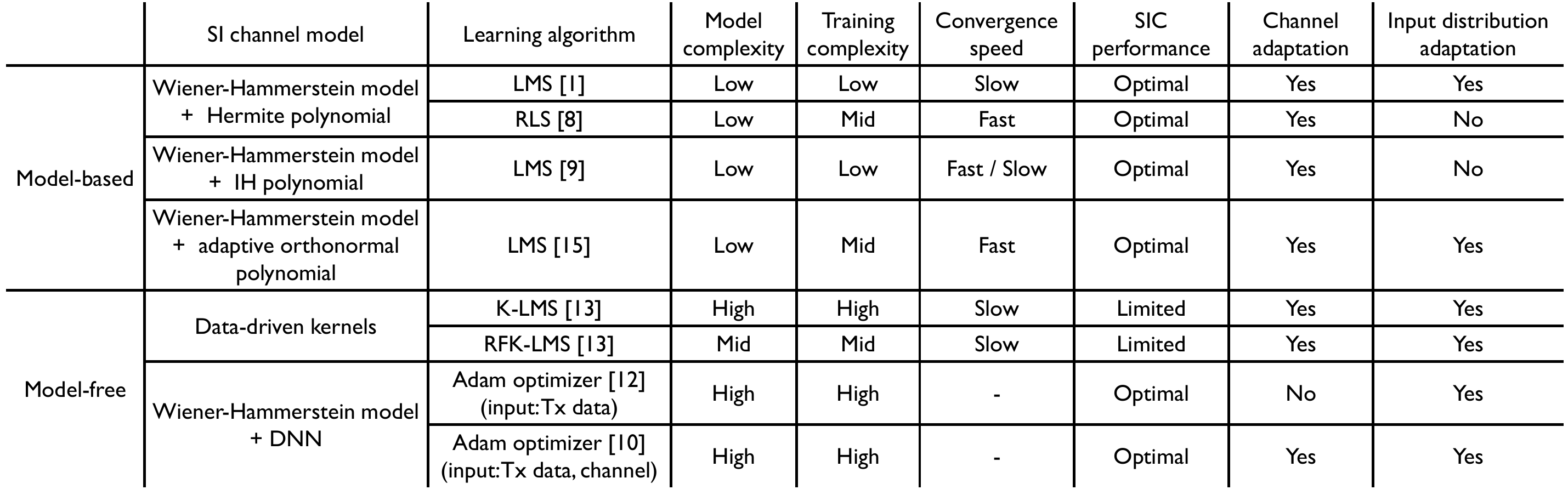}
\caption{A comparison of the model-based and model-free digital SIC algorithms in terms of the model/training complexity, the SIC performance, and the robustness to the change of the channel/data distribution.}
\label{table}
\end{table*}

\subsection{WIH Model with LMS Optimizer}
The Wiener-It\^o-Hermite (WIH) model is an alternative method that uses It\^o-Hermite polynomials (IHPs) instead of HPs to represent the SI channel. The IHPs have the intriguing property that the transformed signals by the IHPs are statistically orthogonal, provided that the transmit data follows a complex Gaussian distribution. This property is advantageous for full-duplex communication systems that use orthogonal frequency division multiplexing (OFDM) because the central limit theorem ensures that the distribution of OFDM converges to a complex Gaussian distribution for a large enough fast Fourier transform (FFT) size.

The orthogonal property of the WIH model allows for the use of a simple LMS optimizer, which achieves the optimal convergence rate of SIC performance for full-duplex systems while having low computational complexity \cite{kim2018adaptive}. However, when the input signal distribution does not follow a complex Gaussian distribution, such as in the case of a single-carrier (SC) waveform with conventional quadrature amplitude modulation (QAM) symbols, the SIC algorithm using the WIH model combined with LMS optimizer slows down the convergence significantly because the orthogonal structure of the IHPs is destroyed.

\subsection{Adaptive WH Model with LMS Optimizer}
 SI canceller that employs the WIH-LMS algorithm faces a major obstacle in that the algorithm requires the input to follow a complex Gaussian distribution. While this is generally true in an OFDM system, it may not be accurate for single-carrier systems where the input distribution may vary due to adaptive modulation and coding on a packet-to-packet basis. To address this issue, the adaptive orthonormal polynomial LMS (AOP-LMS) algorithm has been recently proposed  \cite{lee2023adaptive}. This algorithm is a powerful extension of the WIH-LMS algorithm, designed to tackle non-stationary and arbitrary transmit data distributions. Unlike the WIH-LMS algorithm, which uses the HPs as the basis functions, the AOP-LMS algorithm generates a set of orthonormal basis functions using the moments of the input distribution. This makes it better suited to handle input distributions that may change over time. To estimate the moment parameter, the AOP-LMS algorithm uses a sample mean estimator, which requires a small number of training samples. The orthonormal property of the basis functions enables the simple LMS optimizer to achieve the optimal convergence rate for SIC in full-duplex systems while maintaining low computational complexity. Thanks to these advantages, the AOP-LMS algorithm is a highly effective tool for handling non-stationary and arbitrary input distributions in a variety of applications.

\section{Data-Driven Digital SIC}\label{sec3}
In this section, we review recent advancements in model-free SIC algorithms that leverage machine learning techniques. The popularity of model-free approaches for digital SIC is on the rise due to the increasing abundance of datasets and the enhanced power of modern deep learning pipelines. The primary objective of model-free SIC techniques is to learn the expected output from a vast amount of input data. In contrast to model-based approaches that leverage domain knowledge for the nonlinear SI channel, model-free frameworks treat the nonlinear SIC channel as a black box and train their parameters using input and output data points. The key factor in model-free SIC algorithms is their ability to train the black box accurately. However, the major challenge is that the algorithm and the trained function should be adaptable to track the time-varying channel of SI. Overall, the success of the model-free SIC approach lies in its capacity to learn effectively from data and adapt to the evolving SI channels.

\subsection{Data-Driven Kernel Model with LMS Optimizer}
A data-driven approach to modeling the nonlinear SI channel is to use kernel-based methods in reproducing kernel Hilbert spaces (RKHSs) \cite{ attar2022parallel}. In statistical learning theory, an unknown function defined over RKHS can be represented as a finite linear combination of kernel products evaluated on the input points in the training set data. Selecting an appropriate reproducing kernel function with continuous, symmetric, and positive definite properties is crucial. The Gaussian kernel is a commonly used choice for kernel adaptive filters due to its universal modeling capability, desirable smoothness, and numerical stability. Its properties make it a suitable choice for representing the nonlinear SI channel using data-driven methods.

The kernel least-mean-square (K-LMS) algorithm is a modified version of the LMS algorithm, which operates on RKHS. While the traditional LMS algorithm operates on a finite-dimensional vector space, K-LMS operates on an infinite-dimensional space, which makes it more effective in nonlinear signal processing tasks. However, the na\"{\i}ve K-LMS method suffers from a major drawback - it requires the optimization of an enormous number of parameters. This is because the algorithm expresses the parameters as a linear combination of all previous and current input data, each weighted by their corresponding a priori errors. As a result, storing all past data requires increasing amounts of system memory, as shown in the block diagram of K-LMS in Fig. \ref{fig: SIC}, making it impractical for real-time implementation.

\begin{figure*}
\centering
\includegraphics[width=0.9\linewidth]{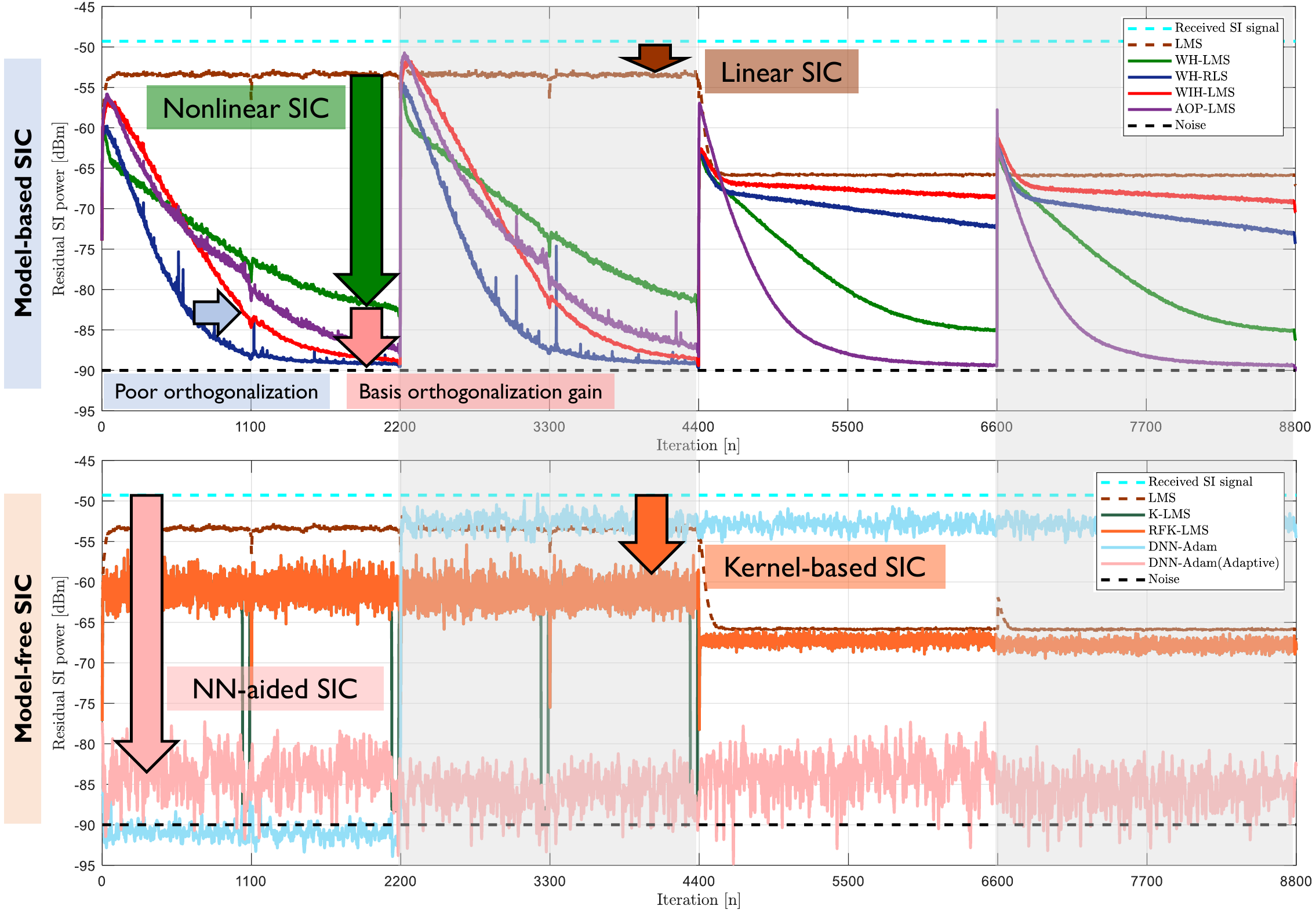}
\caption{The residual SI power comparison according to the various SIC methods. }
\label{fig: res}
\end{figure*}

Traditionally, the kernel trick in RKHS has been used to implicitly map data into a feature space. However, recent developments have seen the adoption of random featured kernel (RFK) nonlinear maps that explicitly map input data to a finite low-dimensional Euclidean space. RFK enables the creation of kernel adaptive filters that use a fixed number of trainable parameters, making them hardware-friendly. One such algorithm, the RFK-LMS, was proposed in \cite{ attar2022parallel}. Although it uses a finite number of trainable parameters, the number required to accurately represent a non-linear function is generally unknown. Therefore, selecting an insufficient number of trainable parameters could lead to failure in accurately representing the unknown function.

Despite its potential benefits, no digital SIC algorithm using RKF-LMS has been proposed in the context of full-duplex systems. We implemented this algorithm for digital SIC and compared its performance with the model-based methods, which will be discussed in Section V.

\subsection{Deep Neural Networks with Adam Optimizer}
DNNs are highly flexible architectures that have shown great promise in modeling nonlinear SI channels with trainable parameters. This makes them an attractive alternative for modeling the nonlinear SI channel without any prior knowledge of the PA nonlinearity \cite{Alexios2018neural}. Notwithstanding this model-free approach, however, DNNs require an offline training procedure, meaning that their parameters need to be pre-trained before performing SIC. Moreover, they need to update the parameters of the DNN according to time-variations in the channel and the transmit data distribution. Unfortunately, implementing sample-by-sample updates of the DNN parameters can be challenging, which limits the applicability of adaptive nonlinear filters with sequential inputs. Therefore, although DNNs are a powerful tool for modeling the SI channel, addressing their offline training requirement and limited applicability in time-varying and non-stationary transmit data scenarios is crucial to fully leverage their potential in practical systems. 

In a recent study \cite{kong2022neural}, a promising approach that combines DNNs and the adaptive method was proposed. This approach uses DNNs to model the nonlinear distortion introduced by the PA, while the linear channel estimator helps to track channel variations. More specifically, the approach uses not only the transmit data but also the estimated linear channel information as the input of the network to track the channel variation. However, a common drawback of DNN-based SIC algorithms is their huge complexity of the offline training process. In such cases, the DNN parameters need to be trained by numerous cases to incorporate the variations in the channel, which hinders the practical real-time full-duplex radio implementation under the non-stationary scenario.

\section{Performance Evaluation} \label{sec4}
In this section, we provide a SIC performance comparison for model-based and model-free algorithms discussed in Sections \ref{sec2} and \ref{sec3}. From the comparison, we provide a discussion on the existing digital SIC algorithms by capitalizing on important implementation aspects, including model complexity, training complexity, convergence speed, SIC performance, channel adaptation, and the non-stationary distribution of transmit signals like in Table \ref{table}.  

\subsection{Simulation Setups}  
In our simulations, the transmitter considers both OFDM and SC transmissions with quadrature amplitude modulation (QAM) symbols. A transmit power is set to 20 dBm, while the received SI signal power after the analog SIC is assumed to be about -50 dBm, and the noise power is -90 dBm. The nonlinearity of the system is designed using the PA modeling, as described in \cite{ding2004digital}. To demonstrate the efficacy of robustness in dealing with variations in the channel and transmit data distributions, we conducted an experiment where a transmitter sent the first 4400 symbols using OFDM, followed by transmitting 1024-QAM symbols for the remaining duration. The channel varies every 2200 symbols.

\begin{figure}
\centering
\includegraphics[width=0.9\linewidth]{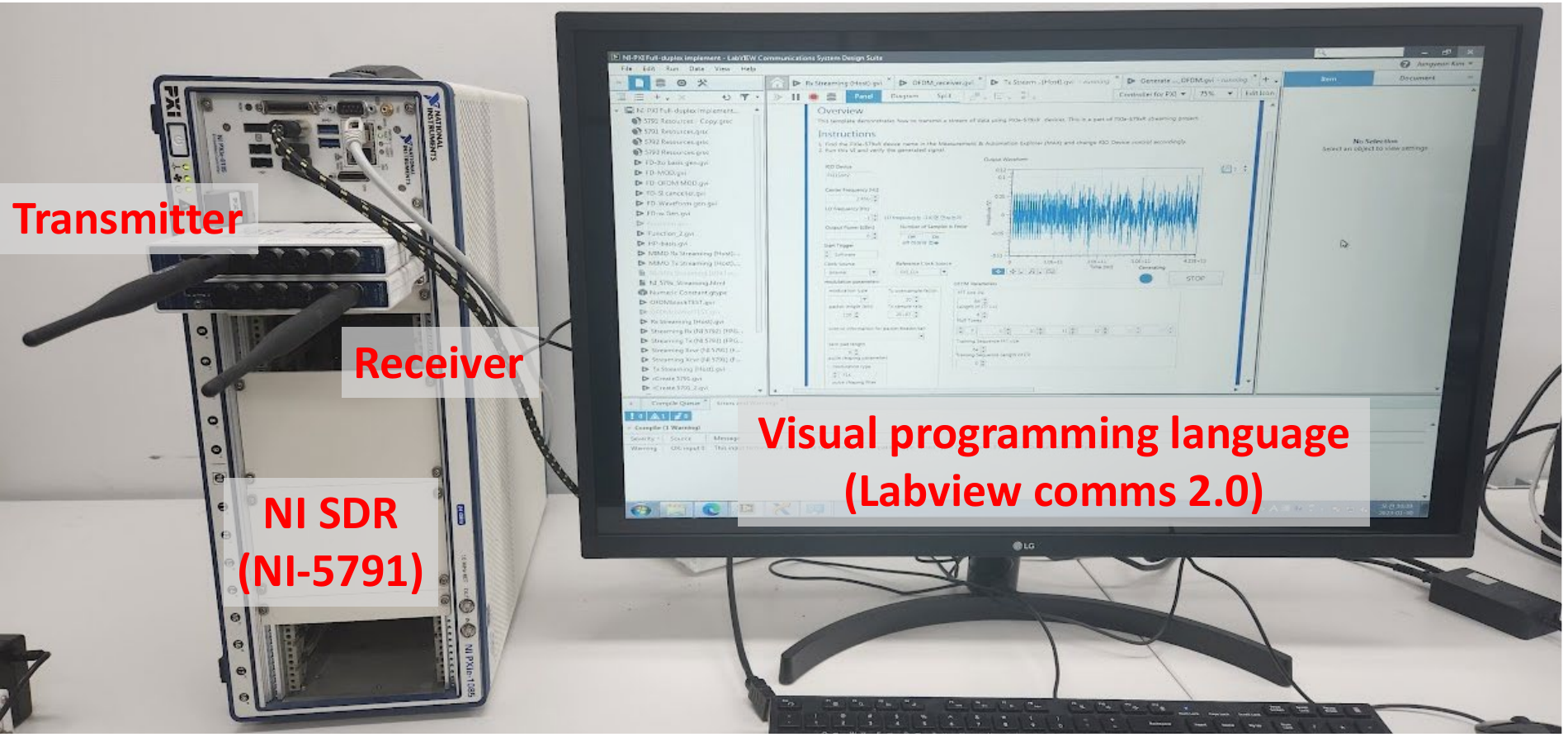}
\caption{A snapshot of the real-time implementation testbed with the NI-PXI SDR platform.}
\label{fig: pxi}
\end{figure}

 \begin{figure*}
\centering
\includegraphics[width=0.9\linewidth]{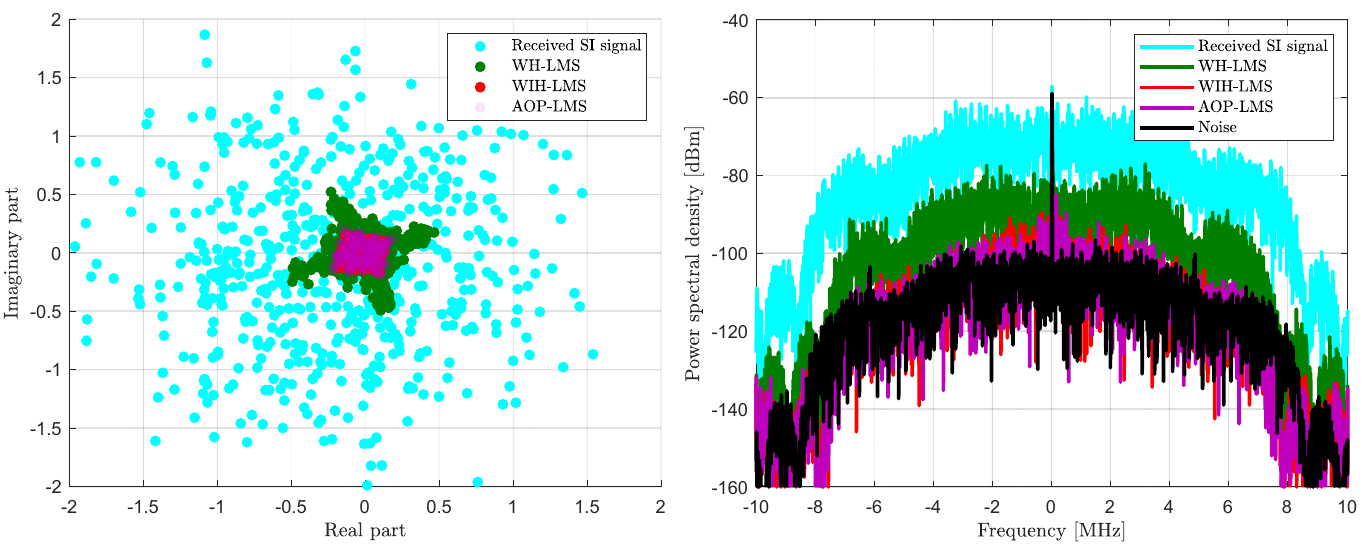}
\caption{Scattering plots and power spectral density of received and the residual SI signals after performing SIC.}
\label{fig: pxires1}
\end{figure*}

\subsection{Complexity Comparison}
\begin{itemize}
    \item {\bf WH-LMS} \cite{bharadia2013full}: The model complexity is measured by the number of filter coefficients used in LMS algorithms. In our WH-LMS implementation,  Within the realm of model-based algorithms,  three concurrent FIR filters are employed, specifically utilizing a 5th order model, each with a length of 21 coefficients. Consequently, the calculated model complexity for this algorithm amounts to 63 coefficients.
    \item {\bf WH-RLS} \cite{Korpi2016Fullduplex}: The WH-RLS algorithm stores the input of the 3 parallel filters for each symbol input and calculates the sample covariance. Then, the eigen-decomposition of the sample covariance is harnessed for the input orthogonalization. Therefore, the model complexity of the WH-RLS algorithm is equivalent to that of the WH-LMS algorithm, but additional computations are required in each iteration compared to the WH-LMS algorithm.
    \item {\bf WIH-LMS} \cite{kim2018adaptive}: The WIH-LMS algorithm assumes the complex Gaussian input. As no additional computations are required for the input orthogonalization, the WIH-LMS algorithm operates with the same model complexity as the WH-LMS algorithm.
    \item {\bf AOP-LMS} \cite{lee2023adaptive}: The AOP-LMS algorithm stores the input in the same manner as the WH-RLS algorithm, but it doesn't require the eigen-decomposition for input orthogonalization. Thanks to the low-complexity Gram-Schmidt-based input orthogonalization, the additional computation complexity was reduced compared to the WH-RLS algorithm. 
    \item {\bf K-LMS} \cite{ attar2022parallel}: The number of filter coefficients to be estimated within the kernel-based algorithm corresponds directly to the count of kernels employed. Given that the K-LMS algorithm evaluates the latest input against all historical inputs through a Gaussian kernel function, its operation encompasses a total of 8800 kernels. Consequently, the K-LMS algorithm exhibits a model complexity amounting to 8800. Moreover, as new inputs emanate from the kernel function, drawing upon both preceding and current data, the process necessitates supplementary computations and memory resources to generate these novel inputs.

    \item {\bf RFK-LMS} \cite{ attar2022parallel}: Rather than comparing the new input with previous inputs, the RFK-LMS employs a distinctive approach, evaluating them against a predetermined collection of randomly generated data through a Gaussian kernel function. With a dataset comprising a size of 500, the RFK-LMS algorithm's model complexity becomes inherently associated with this quantity.

    \item {\bf DNN with Adam optimizer} \cite{Alexios2018neural}: Due to the model-based algorithm's filter length being set at 21, the DNN established an equivalent input node count of 21 to ensure a balanced comparison. When interfacing with the real-number-focused Adam optimizer, this count doubled to 42. The architecture comprises a solitary hidden layer boasting a depth of 1, housing a substantial 200 nodes within. The task involves the estimation of a total of 9002 coefficients, encompassing 8800 weights and 202 biases, thus culminating in a DNN model complexity of 9002. This intricately designed network underwent training over the course of 30000 epochs, employing a diverse training dataset of 2200 instances generated at random.

    \item {\bf Adaptive DNN with Adam optimizer} \cite{kong2022neural}: To enhance adaptability within the Deep Neural Network (DNN), we integrate channel values into the network's input dataset. Consequently, the DNN's input encompasses both the input data and its corresponding channel pair, resulting in a total of 84 input nodes. Recognizing the heightened diversity in inputs, we augment the number of hidden nodes to 300. The adaptive DNN's model complexity is quantified at 26,102 parameters, comprising 25,800 weights and 302 biases. Through a training process spanning 40,000 epochs, we fine-tune this network using a training dataset of 8,800 instances, randomly generated for comprehensive coverage.

\end{itemize}

\subsection{Performance Comparison}

Fig. \ref{fig: res} clearly illustrates the limitations of model-free approaches in canceling nonlinear SI signals. The kernel-based algorithms exhibit a subpar attenuation of approximately -10 dB to -20 dB. The DNN-aided SIC technique shows good SIC performance, but it cannot track the channel varying. Even though the adaptive DNN-aided SIC technique employing the Adam optimizer showcases the effective SIC among the model-free frameworks, its SIC performance was inferior to the mode-based algorithms and it carries a significantly higher complexity and offline training process. Conversely, the WH-LMS algorithm achieves an impressive attenuation of over -35 dB due to its ability to match the nonlinear PA model. However, it suffers from a slow convergence rate. To address this issue, the WH-RLS, WIH-LMS, and AOP-LMS algorithms incorporate orthogonalization gains in their basis functions, resulting in a rapid convergence rate and achieving noise-level SIC power within approximately 2000 iterations. While the WIH-LMS algorithm exhibits a slightly slower convergence rate due to a distribution mismatch between the Gaussian signal and the OFDM modulated signal, its low complexity remains acceptable. Among the algorithms, only the AOP-LMS algorithm maintains noise-level SIC even after variations in the input distribution. Notably, the AOP-LMS algorithm stands out as it attains optimal SIC power, nearly optimal convergence rate, and robustness to channel and input distribution variations in a full-duplex system, all while maintaining reasonable computational complexity. Therefore, it is prudent to leverage model-based approaches when prior knowledge of the system model is available and the model mismatch effect is not pronounced, as model-free approaches may not yield favorable outcomes in such scenarios.

\section{Real-Time Implementation} \label{sec5}

We implemented a full-duplex testbed, adhering to the IEEE 802.11a standard. The implementation operates at a center frequency of 2.4 GHz, with a system bandwidth of 10 MHz. The transmit symbols are generated using an OFDM waveform. For this testbed, both the transmitter and receiver are equipped with omni-directional single antennas, while the noise floor level at the baseband of the system is approximately -100 dBm. In this testbed, we implemented the model-based algorithms for SIC without knowing any true PA model to validate the effect of the model-based approach when the PA modeling error exists. For more information on the experimental results, we refer to our website {\em https://wireless-x.korea.ac.kr/full-duplex-radios}.

Fig. \ref{fig: pxires1} presents a snapshot of the received SI signal power and the corresponding power spectrum density (PSD) for the model-based SIC algorithms. The results clearly demonstrate that the WIH-LMS and AOP-LMS algorithms outperform the WH-LMS algorithm due to their superior model selection that incorporates orthogonality. In such a scenario, we can observe that the WH-LMS algorithm fails to converge adequately while the WIH-LMS and AOP-LMS converge. It is important to note that the AOP-LMS algorithm aligns with the WIH-LMS algorithm when the transmit signal follows a complex Gaussian distribution, such as in the case of OFDM signaling. Our implementation results indicate that both the AOP-LMS and WIH-LMS algorithms can be practical solutions that exhibit the best convergence performance in SIC even when the PA model is unknown.

\section{Conclusion and Discussion} \label{sec7}

This article provided a comparison of state-of-the-art digital SIC algorithms for full-duplex radios by focusing advantages and disadvantages of model-based and data-driven approaches. By conducting a thorough comparative study, we highlighted the numerous benefits associated with adopting a model-based approach for SIC algorithms in full-duplex systems. Utilizing prior knowledge of the system model, a model-based approach demonstrated superior SIC performance while effectively optimizing implementation costs, especially when considering practical scenarios with minimal model mismatch effects. However, in cases where modeling errors are prominent, the data-driven approach emerges as a more favorable option compared to model-based methods. Employing DNNs, the data-driven approach can be an effective strategy for implementing SIC algorithms in full-duplex radios.

To further enhance the advancements in data-driven SIC techniques, particularly in the context of time-varying full-duplex radio systems, the integration of few-shot learning or online learning algorithms can play a pivotal role. These techniques pave the way for the development of more generalized full-duplex communication in future wireless systems.

\bibliographystyle{IEEEtran}
\bibliography{abrv, reference}
{\vskip -1\baselineskip plus -1fil}
\begin{IEEEbiographynophoto}
    {Jungyeon~Kim} is with the Department of Electrical Engineering, POSTECH, Pohang, South Korea.
\end{IEEEbiographynophoto}
{\vskip -1\baselineskip plus -1fil}
\begin{IEEEbiographynophoto}
    {Hyowon~Lee} is with the Department of Electrical Engineering, POSTECH, Pohang, South Korea.
\end{IEEEbiographynophoto}
{\vskip -1\baselineskip plus -1fil}
\begin{IEEEbiographynophoto}
    {Heedong~Do} is with the School of Electrical Engineering, Korea University, Seoul, South Korea.
\end{IEEEbiographynophoto}
{\vskip -1\baselineskip plus -1fil}
\begin{IEEEbiographynophoto}
    {Jinseok~Choi} is with the School of Electrical Engineering, KAIST, Daejeon, South Korea.
\end{IEEEbiographynophoto}
{\vskip -1\baselineskip plus -1fil}
\begin{IEEEbiographynophoto}
    {Jeonghun~Park} is with the School of Electrical and Electronic Engineering, Yonsei University, Seoul, South Korea.
\end{IEEEbiographynophoto}
{\vskip -1\baselineskip plus -1fil}
\begin{IEEEbiographynophoto}
    {Wonjae~Shin} is with the Department of Electrical and Computer Engineering, Ajou University, Suwon, South Korea.
\end{IEEEbiographynophoto}
{\vskip -1\baselineskip plus -1fil}
\begin{IEEEbiographynophoto}
    {Yonina~C.~Eldar} is with the Math and Computer Science Faculty, Weizmann Institute of Science, Rehovot, Israel.
\end{IEEEbiographynophoto}
{\vskip -1\baselineskip plus -1fil}
\begin{IEEEbiographynophoto}
    {Namyoon~Lee} is with the School of Electrical Engineering, Korea University, Seoul, South Korea.
\end{IEEEbiographynophoto}

\end{document}